\documentclass[12pt]{article}

\usepackage{gensymb}

\usepackage{times}
\usepackage{amsmath}
\usepackage{amsfonts}
\usepackage{amssymb}
\usepackage{graphicx}
\usepackage{xcolor}

\title{A Gigapixel Computational Light-Field Camera}

\author{Thomas Gregory${}^{1}$, Matthew P. Edgar${}^{1}$,\\ Graham M. Gibson${}^{1}$, Paul-Antoine Moreau${}^{1\ast}$\\
\\
\normalsize{${}^{1}$School of Physics and Astronomy, University of Glasgow, G12 8QQ, UK}\\
\\
\normalsize{$^\ast$Corresponding author:}\\
\normalsize{Email: paul-antoine.moreau@glasgow.ac.uk}
}

\date{}

\begin{document}

\baselineskip24pt

\maketitle

\begin{abstract}
Light-field cameras allow the acquisition of both the spatial and angular components of the light. This has a wide range of applications from image refocusing to 3D reconstruction of a scene. The conventional way to perform such acquisitions leads to a strong spatio-angular resolution limit. Here we propose a computational version of the light-field camera. We perform a one gigapixel photo-realistic diffraction limited light-field acquisition, that would require the use of a one gigapixel sensor were the acquisition to be performed with a conventional light-field camera. This result is mostly limited by the total acquisition time, as our system could in principle allow $\sim$Terapixel reconstructions to be achieved. The reported result presents many potential advantages, such as the possibility to perform large depth of field light-field acquisitions, realistic refocusing along a very wide range of depths, very high dimensional super-resolved image acquisitions, and large depth of field 3D reconstructions.

\end{abstract}

Since the discovery of quantum ghost imaging~\cite{pittman1995optical} and the realisation that it can be reproduced qualitatively through the use of classical correlations~\cite{shapiro2012physics} and through computational means~\cite{shapiro2008computational,duarte2008single} a number of classical and computational correlation imaging techniques have emerged that can be advantageous to implement in contrast to conventional imaging~\cite{moreau2018ghost,altmann2018quantum}. Among these techniques is the single-pixel camera~\cite{duarte2008single}, where filtering patterns are displayed on a programmable spatial modulator, and the resulting correlation signal is detected by a single pixel photo-sensor. The emergence of this technique led to the realisation that an image can indeed be acquired with a single pixel detector, and can be used in a wide range of applications~\cite{edgar2018principles}. This can present advantages when no spatially resolved detectors are accessible (or are prohibitively expensive), for example when imaging at exotic wavelengths~\cite{moreau2018ghost,edgar2018principles}, or at very precise timing resolution~\cite{howland2013photon,sun2016single}. But if a single pixel detector is in this context sufficient to acquire a conventional image a natural question to ask then is whether similar techniques could also present advantages in conditions where spatially resolved detectors are both accessible and affordable? In other words can the use of multiple-pixels together with such correlation imaging techniques lead to an advantage in terms of extracted information?\\
In two experimental examples the use of a few photo-diodes has been shown to present such an advantage. The first uses four spatially separated photo-diodes to obtain four images of a scene, seen from the exact same point of view but under different illumination conditions leading to a difference in the shading of the scene~\cite{sun20133d}. This allowed 3D reconstructions to be performed using a shape from shading algorithm~\cite{sun20133d}. The second uses three photo-diodes to perform a coloured image reconstruction from the individual red green and blue signals~\cite{welsh2013fast}, this was later modified to add an infrared detector and perform simultaneous real-time visible and infrared video acquisitions~\cite{edgar2015simultaneous}.\\
In other experiments, the use of a camera together with spatial correlations have been achieved to perform ghost acquisitions of a temporal signal. This was done computationally~\cite{devaux2016computational}, with pseudo thermal correlations~\cite{devaux2016temporal} and with quantum correlations~\cite{denis2017temporal}. But the advantage in terms of imaging through combining the use of a correlation imaging technique together with a camera is that techniques that are capable of accessing supplemental spatial quantities of the light-field may be implemented such as in so called light-field imaging~\cite{adelson1992single,ng2005light} (also called plenoptic imaging) or phase-amplitude imaging. It was suggested that schemes with two cameras could harness pseudo-thermal correlations to perform light-field imaging~\cite{pepe2016plenoptic}. The same group was able to implement this proposal experimentally~\cite{pepe2017diffraction}, demonstrating the possibility of obtaining diffraction limited acquisitions through this technique. These concepts were also extended to the use of quantum correlations~\cite{pepe2016correlation}. Additionally, through a similar technique we have recently proposed and experimentally demonstrated the use of quantum correlations, combined with scanning techniques that allows access to supplemental spatial information in order to perform phase-amplitude imaging~\cite{aidukas2019phase}. The main limitation of the aforementioned techniques however, lies in the complexity of the physical systems used to generate the correlations, leading to limited image quality, far from what a photograph could achieve for example.\\
Here we propose and implement experimentally a computational adaptation of these concepts. By using a spatially resolved detector together with a programmable spatial light modulator (SLM), which in our implementation is a Digital Micro-mirror Device (DMD), we acquire photo-realistic and diffraction limited images of a 3-dimensional scene. We demonstrate that the technique can lead to very high dimensional light-field acquisition by performing single acquisitions of 256x256 images (angular resolution) composed of 128x128 pixels (spatial resolution) corresponding therefore to a one Gigapixel light-field acquisition. In fact the technique generally makes use of a camera with $n_c^2$ pixels and a spatial light modulator with $n_s^2$ pixels, that is a total of $n_c^2+n_s^2$ pixels, to lead to a light-field reconstruction with $n_c^2n_s^2$ pixels. This represent a quadratic advantage in contrast to the conventional light-field acquisitions~\cite{adelson1992single,ng2005light} whose total spatial and angular resolution is limited by the total resolution of the camera in use, and which therefore requires a trade-off between the spatial and angular resolution.\\
Light-field imaging has a wide range of applications from photography refocusing~\cite{ng2005light}, 3D reconstructions~\cite{heinze2016automated}, velocimerty and particle tracking~\cite{fahringer2015volumetric,hall2016comparison} to microsopy~\cite{levoy2006light,prevedel2014simultaneous}.\\

\begin{figure}[!t]
\centering
\includegraphics[width=\linewidth]{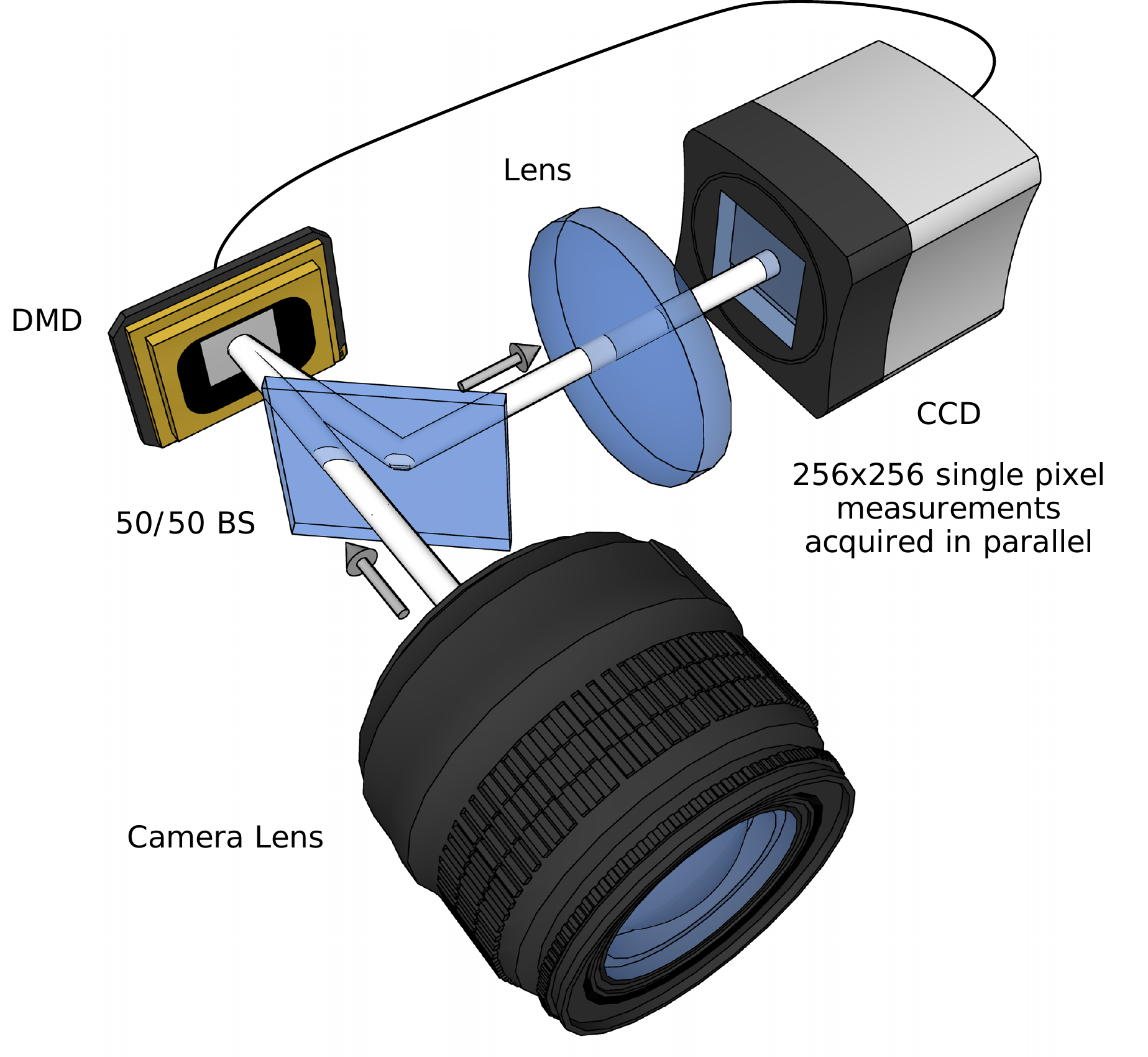}
\caption{\textbf{Computational light-field Imaging setup.} A Camera lens is used to image a scene onto a Digital Micro-mirror Device (DMD), and that can perform a single pixel measurement on each pixels. The light reflected off the DMD pattern is then reflected by a 50/50 beamsplitter (BS) and sent onto a CCD camera set to image the spatial frequencies in the scene. By displaying Hadamard patterns on the DMD, and acquiring camera images for each displayed pattern, one can perform a single-pixel-like light-field.}
\label{fig:Setup}
\end{figure}

In most of these contexts there is an advantage to be able to access a better depth of field, or angular resolution, while preserving the spatial resolution. By allowing the removal of the total light-field resolution limitation of conventional light-field imaging without the necessity of using physical correlations which leads to a complex implementation and noisy results our demonstration allows photo-realistic and high dimensional light-field acquisitions. The technique can present various advantages in different contexts, for example, access to a very large depth of field in photography; access to a very fine angular resolution; the possibility of performing realistic numerical refocusing along a wide range of depths for a single acquisition; the possibility to perform 3D reconstructions over wide ranges and the possibility to perform dramatic super-resolution reconstitution based on micro-scanning reconstruction methods~\cite{superrespatent}. Finally, if the total acquisition time in our implementation is limited by the actual technology used, we predict that the proposed technique could be used to perform video frame rate one Gigapixel light-field acquisitions by using the same DMD and a camera frame-rate ($\sim20kHz$) that is readily attainable with ultra-fast commercially available cameras that possess more than the required 256x256 pixels~\cite{phantom}.

\section*{Results}

\begin{figure}[!t]
\centering
\includegraphics[width=\linewidth]{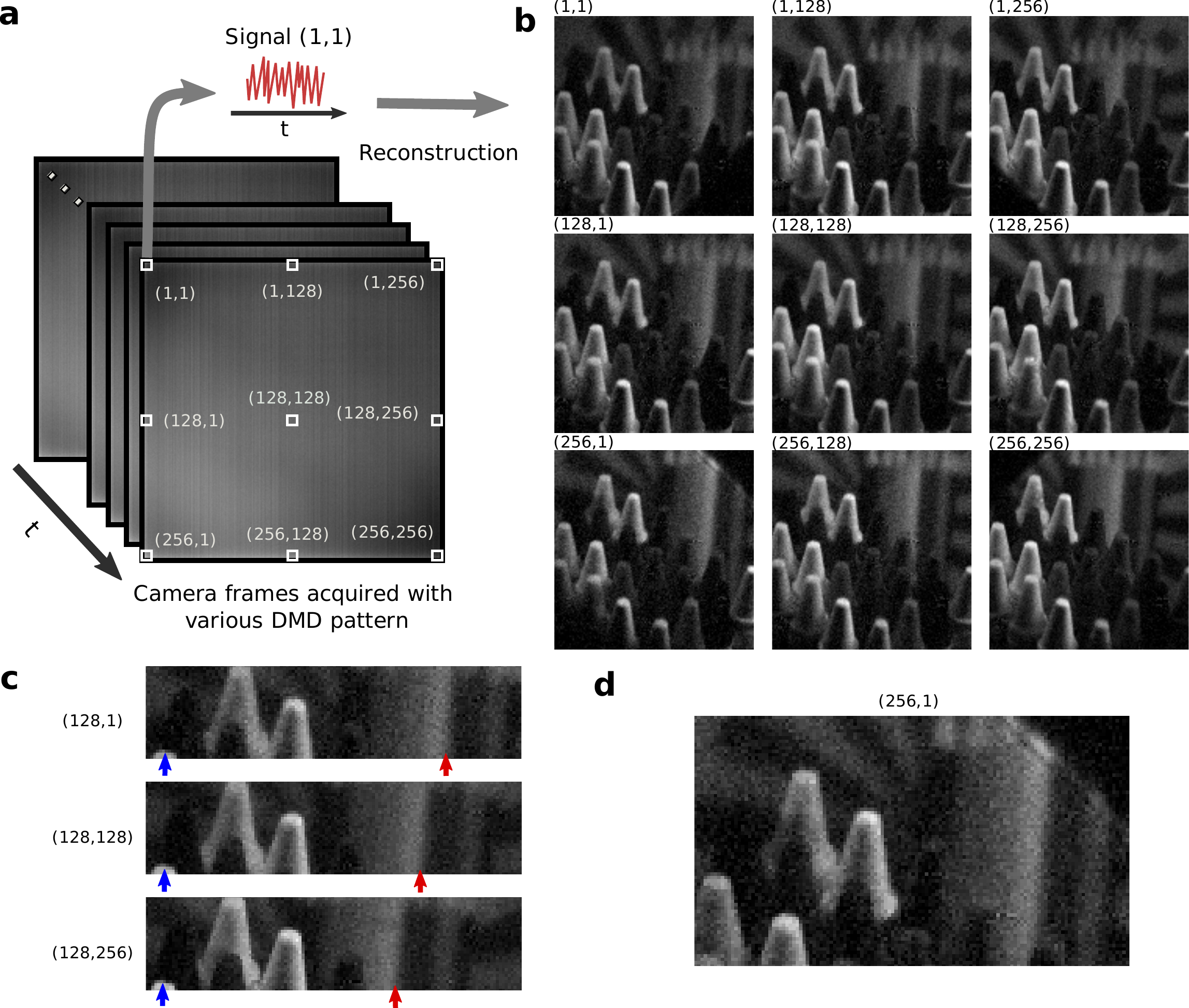}
\caption{\textbf{A computational light-field acquisition.} \textbf(a) Schematic principle of the reconstruction. The set of camera frames acquired with different patterns are used to extract signals. A single signal is obtained by selecting in each of the images a particular $4\times 4$-binned pixel. With each of the $256\times 256$ obtained signals one can reconstruct an image of the scene as seen from a particular POV through a single pixel camera reconstruction. \textbf(b) Nine of the $256\times 256$ reconstructed images obtained for a single light-field acquisition. These images correspond to a reconstruction of the same scene from different points of view, obtained by using camera binned-pixels with coordinate $(x,y)$. \textbf(c) Detail extracted from three reconstructed images. The red and blue arrows highlight the position of two features in the different images. \textbf(d) Detail showing a vignetting effect observable on the images reconstructed using the four pixels in the corner of the CCD camera.}
\label{fig:pov}
\end{figure}

\begin{figure}[!t]
\centering
\includegraphics[width=\linewidth]{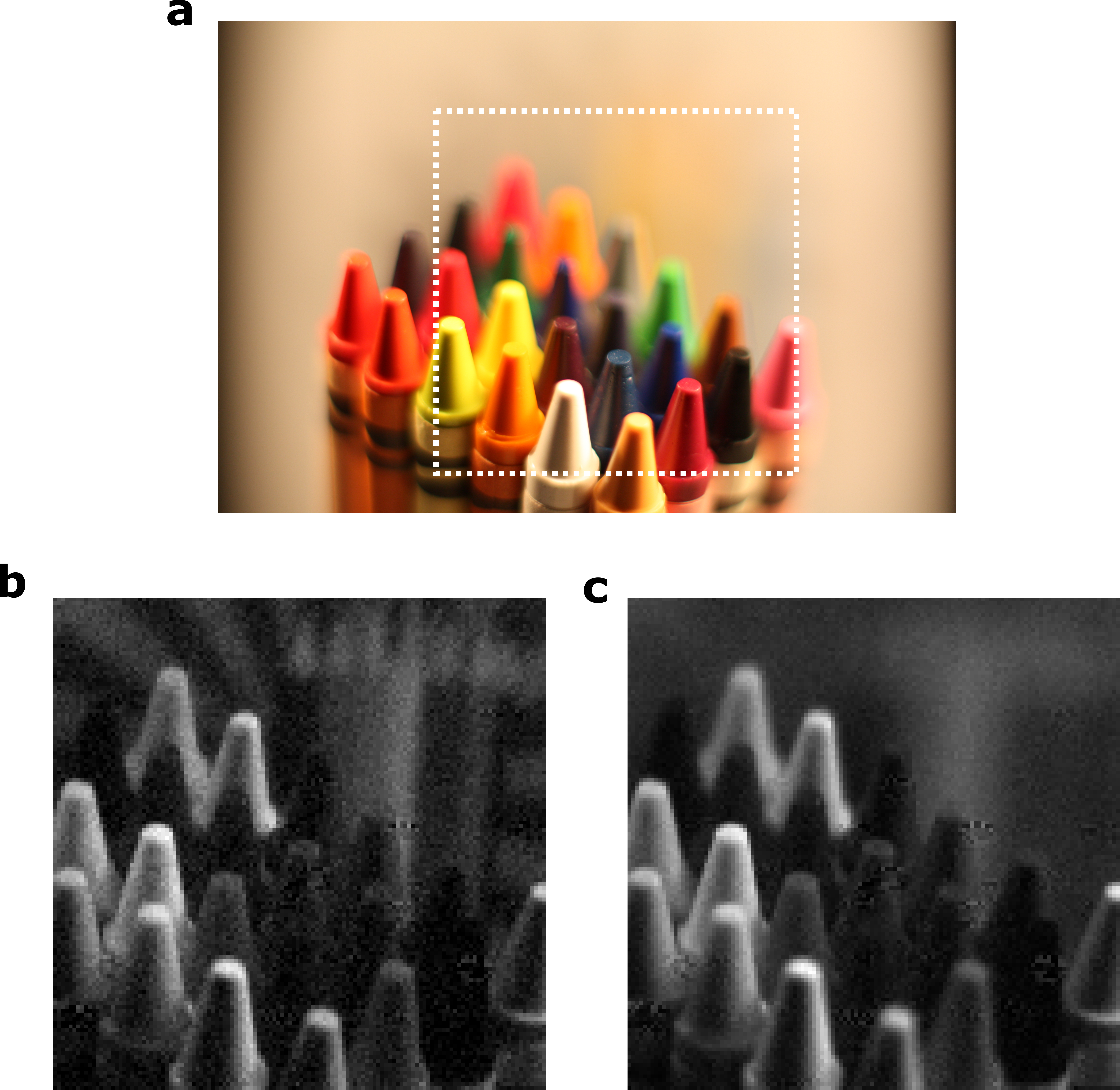}
\caption{\textbf{Comparison of the depth of field in various images. } \textbf(a) Photograph of the same scene obtained with a DSLR camera placed behind the camera lens placed in the same configuration as during the light-field acquisitions. The expected f-number in this context is $\sim1.4$. The dashed line box highlight the field of view obtained in the light-field acquisitions. \textbf(b) Central light-field reconstruction (128,128) using our method. The expected f-number in this context is $\sim1300$. \textbf(c) Integrated single-pixel reconstruction using the same light-field acquisition and the full-frame images intensity as signal to perform the reconstruction. The expected f-number in this context is $\sim5$. }
\label{fig:dof}
\end{figure}

As shown in the setup presented in Fig.\ref{fig:Setup}, the proposed technique consists of using a programmable DMD placed at the back-focal plane of a large aperture camera lens. The DMD operates as a binary device, rapidly actuating each micromirror to one of two states. We can display 2D binary patterns on the DMD and the light reflected along one path off the DMD is collected in an imaging arm by using a 50/50 beamsplitter. The imaging arm is composed of a simple lens and a CCD camera placed in its focal plane. The CCD camera is thus collecting angular information, and the DMD on the other hand can be used to filter spatial information about the scene. The acquisition is then performed by displaying a full set of Hadamard patterns on the DMD and synchronously acquiring images with the camera. The Hadamard pattern set is the set of choice to perform image reconstruction due to their easy mathematical construction and the fact that they form an orthogonal basis~\cite{pratt1969hadamard,souza1988sima}. Using an orthogonal basis avoids redundancy in the acquisition, so that each measurement provides new information about the scene. The patterns are furthermore democratised to avoid prospecting too widely varying signal intensity values, and to stay within the dynamic range of the camera~\cite{sun2017russian}.\\
For each of the camera pixels one can extract a temporal signal and perform a conventional single pixel image reconstruction~\cite{edgar2018principles}. Because the light captured by a particular pixel of the camera corresponds to a particular angle of propagation in the the DMD plane, the scene will ultimately appear as being seen from this particular angle in the reconstructed image. This is much like in conventional light-field imaging~\cite{adelson1992single,ng2005light} where an array of micro-lenses determine the spatial resolution and where the direction of propagation of the light is detected by a subgroup of pixels leading to the reconstruction of a particular point of view. The difference in our implementation here being that each camera pixel is now able to perform this angular measurement for each of the DMD pixels (i.e. for each of the image pixels), that is because through the use of the single-pixel acquisition and reconstruction technique we can untangle the contribution of each of the image pixels within a single CCD camera pixel signal. A schematic of this reconstruction process is shown in fig. \ref{fig:pov}\textbf{a}.\\
In practice we extract 1024x1024 pixel images from the CCD camera and bin the pixels by groups of 4x4 in order to improve the SNR by statistically reducing the relatively high camera noise, for this purpose we also acquire 16 images per Hadamard pattern (128x128 pixels) so as to wash out the effects of the camera noise. By doing so we reconstruct a full light-field acquisition composed of 256x256 points of view of the scene with images composed of 128x128 pixels, that is an effective total acquisition of one Gigapixel. In fig. \ref{fig:pov}\textbf{b} we show nine out of the 256x256 images acquired in a single acquisition. Figure \ref{fig:pov}\textbf{c} illustrates the change of point of view obtained in the resulting images. The three images reconstructed from different binned CCD camera pixels are juxtaposed and the red and blue arrows highlight two particular features in the scene. One can see that depending on the depth of the object compared to the plane of optical focus, the positions of the features are shifted from one image to the other, illustrating the change of angular point of view under which the scene is seen in the different images. Finally in fig. \ref{fig:pov}\textbf{d} we show the presence of a vignetting effect in the images reconstructed with the CCD camera pixels situated in the corner of the sensor.\\
One of the advantages of our technique is that it allows a great depth of field to be accessed, which allows numerical refocusing along a large range of depths to be performed. We show in the method section that in the context of our experiment an f-number of up to $N\sim5000$ can potentially be obtained by using unbinned CCD pixels. Through the reported acquisition, using $4\times 4$ pixel binning, we achieved an f-number of $N\sim 1300$ and when the full CCD sensor signal is integrated over $1024\times 1024$ camera pixels, the obtained effective f-number is $N\sim 5$.\\
In Fig. \ref{fig:dof} we show a comparison of three images of the same scene. Fig. \ref{fig:dof}\textbf{a} is an image acquired with a DSLR camera, Fig. \ref{fig:dof}\textbf{b} is one of the 256x256 images acquired with our system and Fig. \ref{fig:dof}\textbf{c} is an image corresponding to the same light-field acquisition, but for which the signal used to generate the reconstruction was the integrated intensity measured on the whole CCD camera sensor. All three acquisitions were performed using the same camera lens in the same configuration. As expected the depth of field seen on Fig. \ref{fig:dof}\textbf{b}, that is, the depth of field accessible through our light-field reconstruction is greater than the one accessible to a conventional image acquisition with the same lens. Fig. \ref{fig:dof}\textbf{c} illustrates how the depth of field can actually be manipulated by choosing the size of the binned pixels.

\begin{figure}[!t]
\centering
\includegraphics[width=\linewidth]{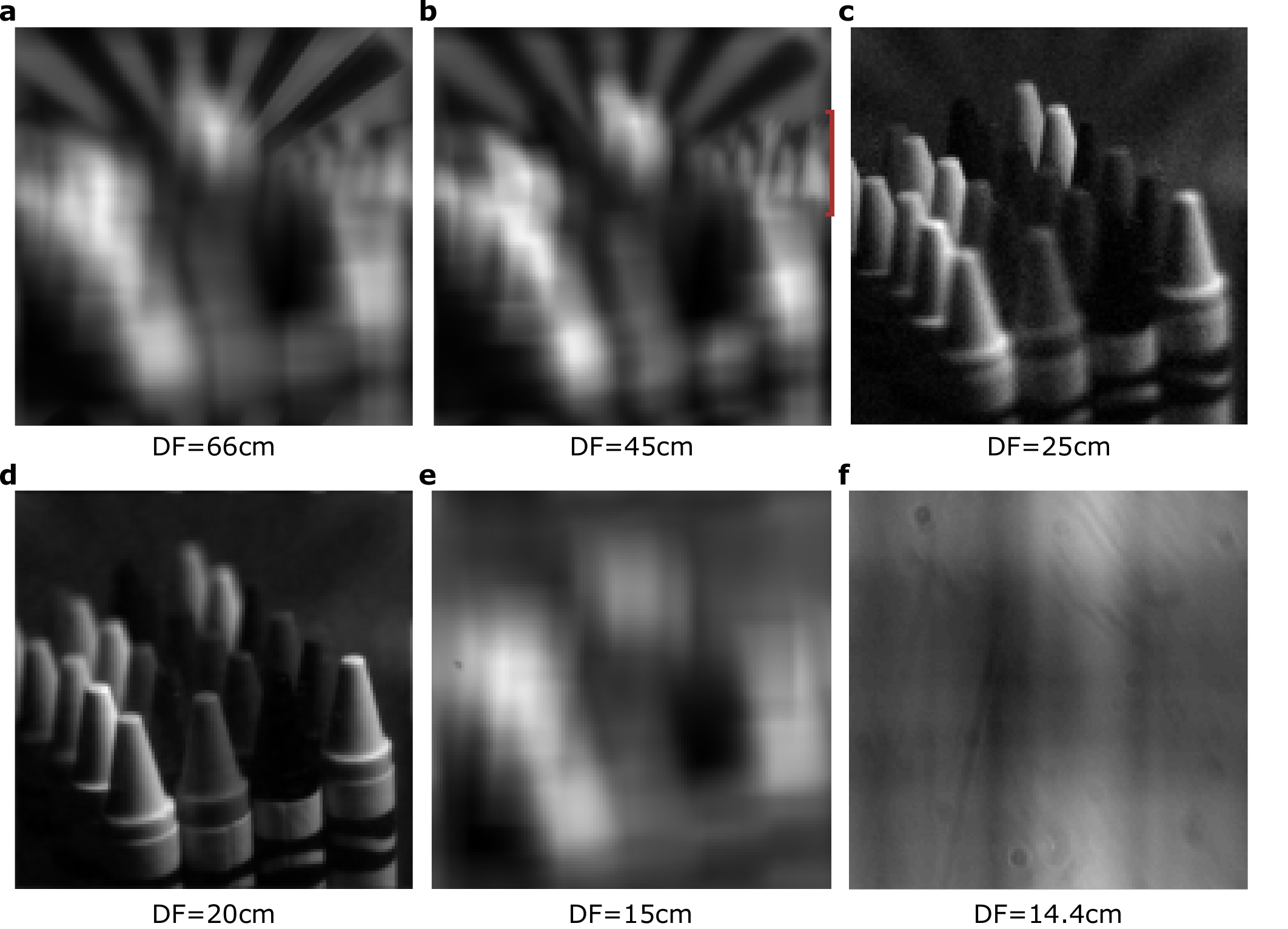}
\caption{\textbf{Refocusing at various depths.} We use a second single computational light-field acquisition of 256x256 POV images of 128x128 pixels to perform a computational refocusing at various depths in the scene. The depth of focus (DF) is the distance between the camera objective and the depth at which we intend to focus in the object space. \textbf(a) Focusing on the spoke target background. \textbf(b) Focusing on the colour pencils (their position is highlighted by a red bracket on the right side of the picture). \textbf(c) Focusing on the farthest crayon. \textbf(d) Focusing on the nearest crayon. \textbf(e) Focusing in front of the scene. \textbf(f) Focusing on dust particles that are not part of the scene but on optical surfaces on the camera lens that are re-imaged on planes close to infinity in the image space.}
\label{fig:refocus}
\end{figure}
We further demonstrate the refocusing capabilities of the light-field acquisitions we obtained. Examples of refocusing at different depths are presented in fig. \ref{fig:refocus}. It can be observed in Fig. \ref{fig:refocus}\textbf{f} that when refocusing close to the object space focal plane, diffraction limited dust particles can be observed on the images, with a clear Airy disk appearance. Indeed, the great numerical apertures accessible, through computational light-field imaging, means that the factor preventing focusing closer and closer to the object focal plane will be the then greater and greater magnification of the objects leading to the images being diffraction limited, thus reaching a fundamental limit similar to the one reported in the physical correlation technique presented in ref.~\cite{pepe2017diffraction}. The observation of such diffraction limited spots is an illustration that the depth of field accessible through the computational light-field imaging that we report allows imaging of optical planes where the resolution limit is no longer bounded by the sensor properties but by the fundamental diffraction limit. This is demonstrated by the fact that the Airy disk pattern is apparent and not blurred by a de-focus effect, i.e. by signal integration over the binned pixel, and also that such diffraction limited spots are not present in images acquired with a conventional camera Fig. \ref{fig:dof}\textbf{a}.

\section*{Discussion}

As shown above the developed computational light-field camera allows the acquisition of photo-realistic diffraction limited images with a very high spatio-angular resolution. By allowing higher angular resolutions to be accessed compared to a conventional light-field camera the technique can present a number of advantages. We list a few in the following.\\
First, the technique presents a great potential for super-resolution based on microscanning reconstruction methods~\cite{superrespatent}. Because an object at a certain depth moves from one point of view to the next relative to the pixel mesh, this can be harnessed to perform a microscanning reconstruction. For a one Gigapixel light-field acquisition one can potentially reconstruct a one Gigapixel image~\cite{superrespatent} as long as the scene is flat enough and out of focus by the right amount so that from one reconstructed point of view to the next the object appears to move by a fraction of the number of angular acquisitions within one pixel. In fact for applications that do not require a speedy acquisition, given the $>1$ Mega pixel camera and the $\sim1$ Mega pixel DMD used here one could in theory reach a $\sim$Terapixel image reconstruction. Obviously performing such drastic super-resolved reconstructions would necessitate solving significant engineering issues, first of all the noise in the images would have to be reduced for the super-resolution reconstruction to be performed. Moreover, the DMD spatial structure, which presents gaps between the individual micromirrors, could prevent the super-resolution reconstruction from performing correctly at sub-DMD-pixel scales. To solve this problem the DMD micro-mirror fill-factor would certainly have to be improved, or the reconstruction method would have to be adapted.\\
A second advantage of the technique lies in the great depth of field that it allows access to. This enables very realistic refocusing along a large range of depth. Obviously, to refocus on a certain plane far from the optical focus point, it needs to appear within the depth of focus in the light-field images in the first instance, but also how realistic the background of a refocused scene is depends on the accessible angular resolution: the more the images integrate a range of points of view (because the angular resolution is not good enough) the more the refocusing will appear non-realistic. That is because in such circumstances, the different points of view integrated in a single light-field image would actually require translating by various amounts in order to perform the refocusing. By removing the spatio-angular resolution trade-off, the computational light-field camera can resolve such issues.\\
A third advantage of our technique can be found in the continuity of the views compared to microlens light-field acquisitions makes the light-field acquisition in itself more realistic. This enables the possibility to navigate realistically within the light-field space.\\
Finally, in the context of 3D reconstruction, the use of a large depth of field and the associated high angular resolution can also present great interest. Using the light-field data extracted from our design to perform multi points of view 3D reconstructions would lead to limited z resolution at large distances due to the fact that the two extreme points of view will be ultimately limited by the camera lens numerical aperture. Nevertheless, the technique presented here has a clear advantage at very short and mid-range distances: the multiplicity of the different points of view represents as many planes to perform a 3D reconstruction along various different depths. Using the full light-field acquisition one could therefore perform 3D reconstructions of a scene on a continuum of short and mid depth ranges, let alone the possibility to use an additional device to also reach long distance ranges.\\

On the other hand, one of the drawbacks of the technique is the speed of the acquisitions. In our demonstration, we were limited to acquire CCD images at a rate around 50Hz due to the limited camera frame rate and the intensity fluctuations of the incandescent light in use that followed the electrical fluctuations of the power supply network. Moreover the large noise of our CCD camera led us to acquire 16 images for each pattern displayed on the DMD, additionally for each pattern we were also displaying its negative, in order to perform a differential measurement and remove further light intensity fluctuations~\cite{ferri2010differential,sun2012normalized}. For these reasons the total acquisition time for a single one Gigapixel acquisition was of the order of around 3 hours ($128\times 128\times 2\times 16=524288$ images at 50Hz). But it needs to be noted that the DMD in use could in theory be used at 20kHz, moreover it has been shown that $\sim 700$ displayed patterns are sufficient to perform 128x128 pixel image acquisitions~\cite{higham2018deep}, and commercially available high speed cameras can reach the DMD speed of 20kHz in terms of frame rate at $>256\times 256$ pixels~\cite{phantom}. With such speeds and techniques, the same one Gigapixel acquisition we report here could potentially be performed at a video frame rate of $\sim 30Hz$.\\
Additionally, the limited spatial resolution reported here is not necessarily a hard limit. If it is true that reaching a spatial resolution higher than $128\times 128$ pixels would slow down the acquisition in the current configuration, one could invert the roles of the two spatially resolved components, i.e. using the CCD camera to detect the spatial components of the field and using the spatial light modulator to measure the angular component of the field, and still perform the full light-field reconstruction through a single pixel reconstruction method. In this way, the time limiting factor would be the angular resolution and not the spatial resolution of the images.\\
Finally, as mentioned above, even though the reported experiment is limited to Gigapixel light-field acquisitions mostly to avoid unbearably long acquisition times, it is to be noted that the spatial resolution of the camera and DMD used here could in principle allow $\sim$Terapixel light-field acquisitions ($1280\times 1024\times 1024\times 768$) to be performed. Such an acquisition would be out of reach of a conventional light-field scheme which would require a Terapixel CCD or CMOS sensor. One can hope that the future improvement in camera and DMD frame rates will allow such acquisitions to be performed in a speedy manner.\\

\section*{Methods}

\subsection*{Materials.} In the experimental implementation of the proposed technique we used a Canon EF50mm f/1.4 USM Camera lens. The single lens placed in front of the CCD camera was a single biconvex 2'' diameter with a focal of f=75mm. The CCD camera used for the angular images acquisition was an Optronis CL600x2/M, with a resolution of $1280\times 1024$ presenting a pixel size of 14 $\mu m$. Finally the DMD used was a ViALUX High Speed V Module (V-7000) with a DMD chip number DLP7000 presenting $1024\times 768$ pixels.\\

\subsection*{Single pixel reconstruction.} The single pixel reconstruction used to generate the light field data is performed in the following way. Let $A_{i,k}$ be a sequence of $N_p$ orthonormal pattern pairs, where pixels can have values of $\pm 1$ ($i$ is the pixel  number  and  $k$ is the pattern  number), the corresponding differential signal obtained on pixel $j$ of the CCD camera is $S_{k,j}$. The reconstructed image of the scene for a particular CCD pixel $j$ (determining the Point of View of the reconstruction) is noted $O_{i,j}$ and is simply estimated as
\begin{equation}
O_{i,j}=\frac{1}{N_p}\sum_{k=1}^{N_p}{S_{k,j}A_{i,k}}
\end{equation}
The set of $O_{i,j}$ represents the full light-field reconstruction.\\

\subsection*{Depth of field.} The depth of field ($DOF$) gives an estimation on the range of positions along the optical axis in the object space that will appear in focus in an image. In the case of the present setup this depth of field is increased through the image angular post-selection that is performed when reconstructing the image using a given camera pixel. The pixels will indeed act as a numerical-aperture-limiting pupil placed upstream of the image sensor in a conventional image acquisition. Because in our configuration the camera has many pixels and that the pixel size is small it is expected that the $DOF$ of the reconstructed images will be great, thereby allowing some details to be brought into focus which could not be obtained with a conventional image acquisition or with a conventional light-field acquisition. This is because for a conventional image acquisition or a conventional light-field acquisition the number of camera pixels per microlens limits the number of accessible points of view, imposing upon each pixel to occupy a relatively large area in terms of the different angles it collects. The $DOF$ can be expressed in the following way~\cite{greivenkamp2004field}: 
\begin{equation}
\mathit{DOF}=\frac{2NL_o^2f^2B}{f^4-N^2L_o^2B^2}
\label{eqn:dof}
\end{equation}
Where $L_o$ is the object position relative to the camera lens entrance pupil, f is the focal length of the lens placed in front of the CCD camera, $B$ is the acceptable blur diameter criterion and $N$ is the f-number. One can observe the role of the f-number in eq. \ref{eqn:dof} that shows that the depth of field is increased when the optical system f-number is increased.\\

The f-number is an image-space, infinite-conjugate quantity~\cite{greivenkamp2004field}, that can be linked to the image space numerical aperture $\mathit{NA}_i$ in the following way, under the small angle approximation:
\begin{equation}
    N\approx\frac{1}{2\mathit{NA}_i}
\end{equation}
In our implementation, $\mathit{NA}_i$ will effectively be limited by the single pixel angular post selection. Let $s_p$ be the size of the effective camera pixels used for the reconstruction, and $f_c$ be the focal length of the camera lens. In the image space, the pixel will act as the angular acceptance pupil and the associated numerical acceptance will be, under the small angle approximation:
\begin{equation}
    \mathit{NA}_i\approx\frac{s_p}{2f_c}
\end{equation}
This leads to the following expression for the effective f-number for the 'single pixel' reconstructed images:
\begin{equation}
    N\approx\frac{f_c}{s_p}
\end{equation}

\subsection*{Refocusing.} 
The refocusing is performed by treating the full one Gigapixel data, simply through a translation of each of the images in each of the two dimensions, by an amount proportional to the coordinates of the binned camera pixels used to reconstruct each image. The translated images are then summed altogether. One can set the value for the smallest translation i.e the difference in image translation from one binned camera pixel to the next. The different values will correspond to different depths of re-focus. The larger this parameter the farther from the optical focus the numerical focus will be.

\subsection*{Presence of noise and 3D stereoscopic viewing.} In Fig. \ref{fig:refocus}\textbf{c} one can observe the presence of noise that is not present in the other refocused images. The depth of focus used in this context is the actual optical focus plane of the camera lens, so that the different images are here simply summed without translation. Because there is a fixed pattern noise shared by all the images, this adds constructively in this image and is washed out in the others because the fixed pattern is then translated independently in the different images before they are summed. One can see this effect as well when observing stereoscopic images using two POV images extracted from our acquisitions. We present an anaglyph 3D in the supplementary text, and also give the two POV images used that can be displayed on a 3D screen or headset (see supplementary text). When observing this stereoscopic image one can see the depth observation 3D effect which shows that our images are candidates to perform 3D stereoscopic reconstructions. Moreover the noise can even be located in depth approximately at the central depth of the crayons, this is where the camera lens was set to focus.

\noindent \textbf{Acknowledgements:}
The authors are grateful to Miles Padgett for useful discussions. P-AM acknowledges the support from the European Union Horizon 2020 research and innovation program under the Marie Sklodowska-Curie Action (Individual MSCA Fellowship no. 706410), of the Leverhulme Trust through the Research Project Grant ECF-2018-634 and of the Lord Kelvin / Adam Smith Leadership Fellowship scheme. TG acknowledges the financial support from the UK EPSRC (EP/N509668/1) and  from the Professor Jim Gatheral quantum technology studentship.

\noindent \textbf{Author Contributions} P-AM initiated the project, conceptualised the demonstration and designed the experiment. TG and P-AM conducted the experiment with the support of MPE and GMG. TG, MPE and P-AM developed the acquisition and analysis tools. TG and P-AM interpreted the results. All authors contributed to the manuscript.\\

\clearpage

\end{document}